\documentstyle[psfig]{l-aa}

\begin{document}

\thesaurus{ 06(08.06.2; 08.12.1; 08.12.2) }

\title{ Medium-resolution optical spectroscopy of young stellar and sub-stellar
M-dwarfs in the Cha I dark cloud \thanks{ Based on observations collected at the 
European Southern Observatory 3.5m NTT on La Silla during program 63.L-0023.}}

\author{ R. Neuh\"auser\inst{1} \and F. Comer\'on\inst{2} }

\offprints{R. Neuh\"auser, rne@mpe.mpg.de }

\institute{MPI f\"ur extraterrestrische Physik, Giessenbachstra\ss e 1, D-85740 Garching, Germany
\and European Southern Observatory, Karl-Schwarzschild-Stra\ss e 2, D-85748 Garching, Germany
}

\date {Received 30 June 1999; accepted 21 July 1999 }

\maketitle

\markboth{Neuh\"auser \& Comer\'on: Young M-dwarfs in Cha I}{}

\begin{abstract}

We obtained medium-resolution spectra of the bona-fide brown dwarf Cha~H$\alpha$~1,
the five brown dwarf candidates Cha~H$\alpha$~2 to 6, two additional late M-type 
brown dwarf candidates, all originally selected by H$\alpha$ emission,
and four previously known T~Tauri stars, all located in Chamaeleon I.
The spectral types of our targets range down to M8.
We show their spectra and also list their IJK magnitudes from DENIS.
All objects have radial velocities consistent with kinematic membership 
to Cha I and show Li 6708\AA~absorption. Our Cha I brown dwarf candidates 
with lithium are young or sub-stellar or both. 
Cha~H$\alpha$~1, 3, 6, and 7 are certainly brown dwarfs:
Either they are as old or older than the Pleiades and should have 
burned all their original lithium if they were late M-type stars, or,
if they are younger than the Pleiades, $\le 125$~Myrs, they are
sub-stellar because of the young age and the late spectral types ($\ge$~M7),
according to three different sets of evolutionary tracks and isochrones.
To classify Cha~H$\alpha$~2, 4, 5, and 8 as either stellar or sub-stellar,
evolutionary tracks of higher reliability are needed.


\end{abstract}

\section{Introduction: Young brown dwarfs}

Recently, Comer\'on et al. (1999, henceforth CRN99) performed an 
H$\alpha$ objective-prism survey in the Chamaeleon I dark cloud, a site of on-going 
and/or recent low- and intermediate-mass star formation, and presented six 
new low-mass late-type ($\ge$ M6) objects with H$\alpha$ emission,
called Cha~H$\alpha$ 1 to 6. Their strong H$\alpha$ emission,
if having the same origin as in T~Tauri stars (TTS), is an indicator of youth, 
and therefore of membership to the star forming region at $\sim 160$ pc.
Luminosities obtained from VIJHK photometry made it possible to place the objects in
the H-R diagram 
where, according to different sets of pre-main sequence evolutionary tracks, 
they all lie near or below the hydrogen burning limit. 
Follow-up spectroscopy with high S/N of the lowest-mass object, Cha~H$\alpha$~1, 
confirmed its sub-stellar nature (Neuh\"auser \& Comer\'on 1998, henceforth NC98) 
and allowed an accurate determination of its spectral type to be M7.5-M8. 
Comparision with evolutionary tracks and isochrones yielded a mass of 
$\sim 0.04~M_{\odot}$ and an age of $\sim 1$~Myrs. Cha~H$\alpha$~1 is the 
first brown dwarf (BD) detected in X-rays (NC98).

Young brown dwarfs were also found in other star forming regions, 
namely $\rho$ Oph (Rieke \& Rieke 1990; Comer\'on et al. 1993, 1998; 
Luhman et al. 1997; Wilking et al. 1999), Taurus (Luhman et al. 1998; 
Brice\~no et al. 1998; White et al. 1999; Reid \& Hawley 1999),
and Orion (B\'ejar et al. 1999).

The sub-stellar nature of suspected brown dwarfs can usually be confirmed 
by detection of lithium 6708\AA~absorption, because lithium is burned rapidly
and fully by proton capture in low-mass, i.e. fully convective stars (Magazz\`u
et al. 1993) with masses above $\sim 0.06~M_{\odot}$. This lithium test cannot be
applied to very young objects, which are not old enough, yet, to have burned 
all their initial lithium: for example, objects younger than $\sim 10$ Myr with a 
mass below $\sim 0.3~M_{\odot}$ have burned less than $1\%$ of their original lithium,
and the percentage is even lower for less massive objects of the same age 
(D'Antona \& Mazzitelli 1997, Baraffe et al. 1998, and A. Burrows, private
communication). For this reason, a lithium detection in the spectrum of a 
suspected brown dwarf member of a star forming region does not directly 
confirm its substellar character, but it does support strongly its membership
in the star forming region (the alternative possibility, namely the object 
is evolved and foreground, automatically implies that it is a brown dwarf 
because of the lithium detection). Once such membership and, hence, distance
is established, the mass can be inferred from the spectral type and luminosity 
by comparison to theoretical evolutionary tracks for very low mass objects.

\begin{figure*}
\vbox{\psfig{figure=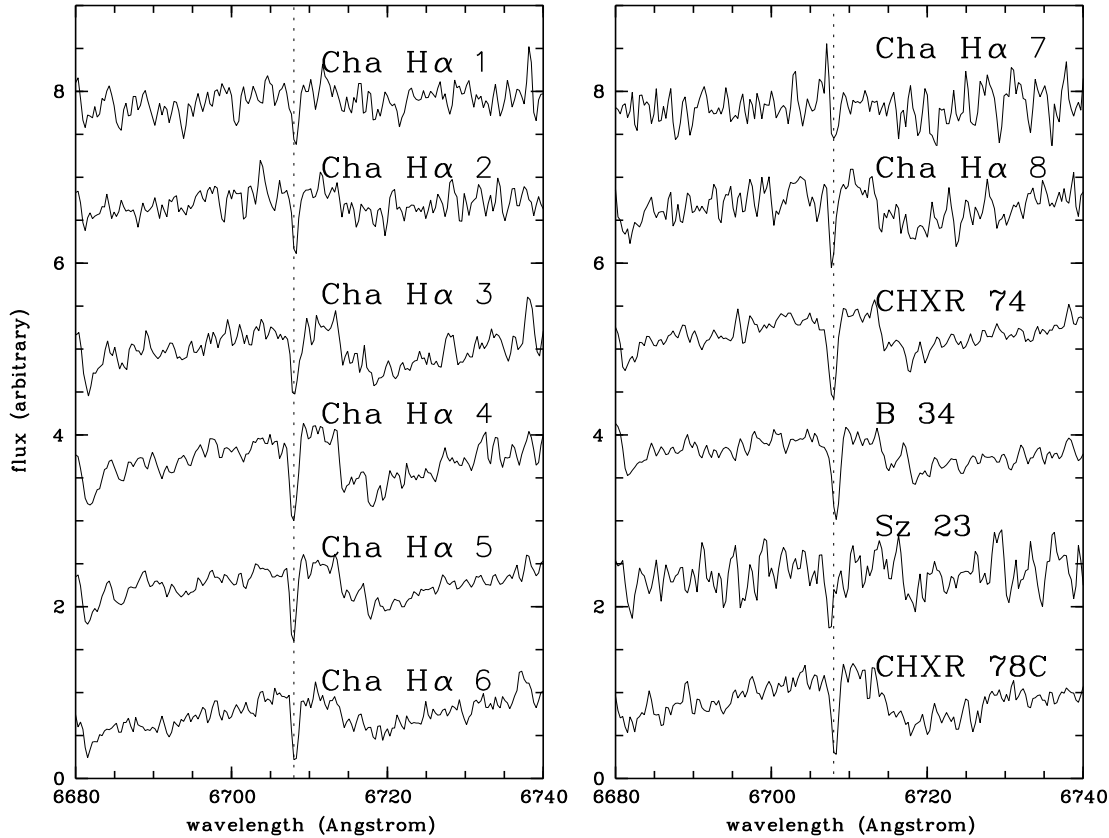,width=18cm,height=12cm,angle=270}}
\caption{ Parts of our medium-resolution spectra around the
lithium 6708\AA~line, which is detected in all objects }
\end{figure*}

In this paper, we present medium-resolution spectra of Cha~H$\alpha$~1 to 6, 
two additional late M-type objects (Cha H$\alpha$ 7 and 8), also first detected 
in an H$\alpha$ objective prism survey (Comer\'on et al., in preparation), 
as well as four M-type TTS in Cha I.
Signal-to-noise ratio (S/N) and resolution of these new spectra are better 
than those in CRN99, so that we can investigate not only H$\alpha$ emission, 
but also lithium absorption and radial velocities. 

\section{Observations and data reduction}

\begin{table*}

\begin{tabular}{lcclccccccccc} \hline
\multicolumn{13}{c}{ \bf Table 1. M-dwarfs targets and results (*)} \\ \hline
Object & \multicolumn{2}{c}{$\alpha, \delta$ (J2000.0)} & spec$^{(1)}$ & 
\hspace{-.5cm} $T_{eff}^{(2)}$ & I & J & K &
\multicolumn{2}{c}{$W_{\lambda}(H\alpha)$~[\AA ]$^{(4)}$} & $W_{\lambda}$(Li) 
& RV$_{lsr}^{(5)}$ & Classification \\
designation & $11h$ & $-77 ^{\circ}$ & type & \hspace{-.5cm} [K] 
& \multicolumn{3}{c}{DENIS$^{(3)}$ [mag]} &
{\scriptsize CRN99} & {\scriptsize this work} & [\AA ] & $[km~s^{-1}]$ & \\ \hline

Cha H$\alpha$ 1 & 07:17.0 & 35:54 & M7.5 & \hspace{-.5cm} 2805 & 16.4 & 13.3 & 12.3 & $-59 $& $-34.5$ & 0.63:& $+0.9 \pm 5.3$ & bona-fide BD \\

Cha H$\alpha$ 2 & 07:43.0 & 33:59 & M6.5 & \hspace{-.5cm} 2940 & 15.3 & 12.1 & 10.6 & $-39 $& $-71.0$ & 0.43 & $+8.4 \pm 1.9$ & candidate BD \\

Cha H$\alpha$ 3 & 07:52.9 & 36:56 & M7   & \hspace{-.5cm} 2890 & 15.0 & 12.3 & 11.1 & $-4.5$& $-14.4$ & 0.43 & $+6.9 \pm 4.9$ & bona-fide BD \\

Cha H$\alpha$ 4 & 08:19.6 & 39:17 & M6   & \hspace{-.5cm} 2990 & 14.4 & 12.0 & 11.1 & $-4.7$& $-9.7 $ & 0.48 & $+6.1 \pm 3.2$ & candidate BD \\

Cha H$\alpha$ 5$^{(6)}$& 08:25.6 & 41:46 & M6 & \hspace{-.5cm} 2990 & 14.7 & 12.0 & 10.7 & $-7.6$& $-8.0 $ & 0.42 & $+4.4 \pm 3.0$ & candidate BD \\

Cha H$\alpha$ 6 & 08:40.2 & 34:17 & M7   & \hspace{-.5cm} 2890 & 15.1 & 12.0 & 10.9 & $-59 $& $-61.7$ & 0.43 & $+0.2 \pm 2.3$ & bona-fide BD \\ 

Cha H$\alpha$ 7$^{(7)}$& 07:38.4 & 35:30 & M8 & \hspace{-.5cm} 2720 & 16.7 & 13.5 & 12.4 & (8) & $-35.0$ & 0.80:& $-1.8 \pm 6.7$ & bona-fide BD \\

Cha H$\alpha$ 8$^{(7)}$& 07:47.8 & 40:08 & M6.5 & \hspace{-.5cm} 2940 & 15.6 & 12.7 & 11.5 & (8) & $-8.4$ & 0.49 & $-2.9 \pm 5.5$ & candidate BD \\ \hline

CHXR 74$^{(9)}$ & 06:57.4 & 42:10 & M4.5 & \hspace{-.5cm} 3200 & 13.7 & 11.6 & 10.2 & $-13$  & $-8.3$  & 0.69 & $+5.9 \pm 1.0$ & T~Tauri star \\ 

B 34 $^{(10)}$ & 07:35.4 & 34:51 & M5   & \hspace{-.5cm} 3125 & 14.4 & 12.1 & 10.9 & $-5.5$ & $-7.2$  & 0.80 & $+7.0 \pm 1.7$ & T~Tauri star \\

Sz 23$^{(11)}$ & 07:59.4 & 42:40 & M2.5 & \hspace{-.5cm} 3500 & 14.6 & 12.0 & 9.9 & $-45$ & $-56.5$ & 0.34 & $-2.8 \pm 6.0$ & T~Tauri star \\

CHXR 78C& 08:54.4 & 32:12 & M5.5 & \hspace{-.5cm} 3060 & 14.8 & 12.2 & 10.9 & $-3.2$ & $-13.4$ & 0.48 & $+9.5 \pm 2.9$ & T~Tauri star \\ \hline

\end{tabular}

\vspace{-.3cm}
(*) Approximate errors are $\pm 0.25$ sub-classes in spectral type, $\pm 50$ K in $T_{eff}$, 
$\pm 0.1$ mag in IJK, $\pm 5$\AA~in $W_{\lambda}(H\alpha)$, and $\pm 0.05$\AA~in $W_{\lambda}$(Li) 
(but $\pm 0.1$\AA~in Cha H$\alpha$ 1 and 7 because of lower S/N). \\
Notes: 
(1) Spectral types (ST) for Cha H$\alpha$ 1 to 8 assigned 
from high-S/N, low-resolution spectra obtained for a much broader wavelength range 
(Comer\'on et al., in preparation) than in Fig. 1; ST for the four TTS
are taken from CRN99.
(2) Effective temperatures $T_{eff}$ converted from ST using Fig. 7 
in Luhman (1999) for objects intermediate between dwarfs and giants, 
as appropriate for young M-dwarfs.
(3) DENIS data for Cha~H$\alpha$~1 to 8 are from L. Cambr\'esy (private 
communication) and for TTS from Cambr\'esy et al. (1998), they all 
compare well with data given in CRN99 and Lawson et al. (1996).
(4) Negative for emission. 
(5) Local standard of rest (lsr) radial velocity (RV).
(6) Coordinates of Cha H$\alpha$ 5 were slightly incorrect in NC98 and CRN99.
(7) For finding charts, see Comer\'on et al., in preparation. 
(8) Not listed in CRN99.
(9) Lawson et al. (1996) detected Li at low S/N and gave $W_{\lambda}(H\alpha) = -2.7$\AA .
Alcal\'a (1994) gave $W_{\lambda}(H\alpha) = -4.1$\AA .
(10) Alcal\'a (1994) gave $W_{\lambda}(H\alpha) = -7.7$\AA~and detected Li.
(11) Double-peaked H$\alpha$ emission, possibly a companion to VW Cha (Reipurth \& Zinnecker 1993).

\end{table*}

We obtained medium-resolution spectra of the Cha I low-mass objects Cha~H$\alpha$~1 
to 8 as well as the TTS CHXR 74, B 34, Sz 23, and CHXR 78C.
The spectra were obtained in the nights of 17 and 18 April 1999 at the 3.5m-New 
Technology Telescope (NTT) of the European Southern Observatory (ESO) on La Silla with 
the ESO Multi Mode Instrument (EMMI) in the red medium dispersion mode.
Grating $\# 6$ was used covering the range 6485 to 7105\AA , 
providing a dispersion of 0.32\AA~per each 0.27" pixel (CCD~$\# 36$), 
and a resolution of $\sim$ 5800 (with the 1.0" slit used).
One or several spectra were obtained for each object, with exposure time and number of spectra 
according to the measured instrumental throughput and the available R-band photometry. To keep 
the cosmic ray density in the resulting images reasonably low, the longest exposure times 
per frame were never longer than 45 min, always placing two objects into the 330" long slit.

Each spectrum was extracted individually from the corresponding bias-subtracted, 
flat-fielded two-dimensional image using tasks in the APEXTRACT package layered on IRAF. 
Wavelength calibration was performed by obtaining spectra of a He-Ar lamp immediately 
before and after each individual exposure, to ensure that wavelength shifts due to the
rotation of the instrument during the exposure in the Nasmyth focus were kept to a minimum. 
A full dispersion solution was obtained for one of the He-Ar spectra, which was then used 
as the master wavelength calibration frame. Zero-point shifts with respect to this spectrum 
were then obtained for each of the other He-Ar lamp exposures and applied to the extracted 
spectra of the targets. These zero-point shifts were based on spectra of the calibration 
lamp extracted over a narrow range of rows of the detector centered on the position of 
the object. In this way, the determined shifts included the corrections due both to 
flexures of the instrument, and to the slight curvature of the spectral lines across the
surface of the detector, which amounts to a significant four-pixel difference between 
the center and the edges. Correction for telluric features was performed by observing 
the bright star $\beta$ Cha at intervals of about 2.5 hours per night. The high rotation 
velocity of this star (255~km/s, Hoffleit et al. 1982) allows an easy separation 
between the narrow telluric and the broad photospheric lines. The spectra of $\beta$ Cha 
were interpolated linearly in the wavelength intervals containing photospheric lines. 
Then, each individual spectrum of our targets was ratioed by the one of $\beta$ Cha 
obtained at the most similar airmass.
The next step was the modification of the wavelength axis of each object to correct 
for the observer's motion with respect to the local standard of rest (lsr).
The lsr velocities for each object were computed with the RVCORRECT task under 
IRAF, and the spectra were then accordingly Doppler-shifted. Finally, all
the wavelength-calibrated, telluric-corrected spectra in the lsr frame for
each object were added together to produce the combined spectra (Fig. 1).
Radial velocities are derived by Fourier cross-correlation using the 
6600 to 7100\AA~part of the spectrum of CHXR 74 as a reference template.
The zero-point is derived from the lithium doublet, 
using 6707.815\AA~as central wavelength. 
The quality and S/N of the spectra were insufficient, 
however, to obtain reliable rotational velocities.

Results are listed in Table 1.
In addition, we obtained ground-based, high-S/N, low-resolution optical 
and infrared spectra over a much broader wavelength range than shown 
in Fig. 1 or in CRN99, in order to better constrain the spectral types
of the Cha H$\alpha$ objects, as well as additional photometry and ISO data,
all of which are to be presented elsewhere (Comer\'on et al., in preparation). 
Spectral types -- as listed in Table 1 -- are determined by comparison with 
late M-type standards (Kirkpatrick et al. 1991, 1995) and spectra indices as in CRN99. 
We converted them to effective temperatures $T_{eff}$ using the 
relation
recently presented by Luhman (1999) for young M-type objects in IC 348.

\section{Results and discussion}

To investigate kinematic membership to Cha I, we compare the radial velocities 
of our objects with those of known TTS members of Cha I and the local molecular gas, 
both having lsr radial velocities peaking at $\sim 4$ km/s with a $1~\sigma$
scatter of $\sim 6$ km/s (Dubath et al. 1996, Covino et al. 1997, see also 
our Fig. 2). The TTS observed here also have such velocities.
Hence, within the errors, Cha~H$\alpha$~1 to 8 all have radial velocities in 
this range, i.e. are kinematic members of the Cha I association (see Fig. 2).

Cha I membership can also be investigated by looking for youth signatures, 
such as H$\alpha$ emission or lithium absorption. Gravity sensitive features 
and thermal infrared excess emission observed by ISOCAM will be discussed 
in a companion paper (Comer\'on et al., in preparation).

\begin{figure}
\vspace{-.3cm}
\vbox{\psfig{figure=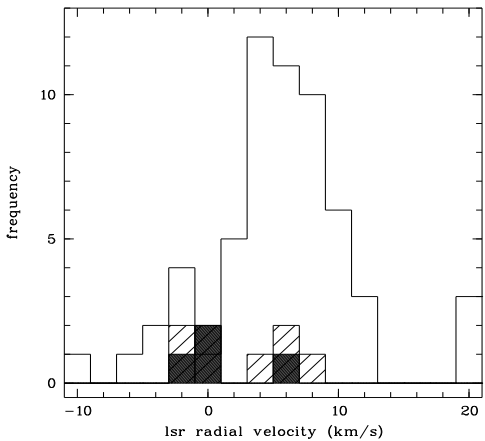,width=11cm,height=7cm,angle=270}}
\caption{ Local standard of rest (lsr) radial velocities of
Cha TTS (Dubath et al. 1996, Covino et al. 1997, and Table 1),
candidate BDs (hatched) and bona-fide BDs (filled) }
\end{figure}

For most of our objects, $W_{\lambda}(H\alpha)$ is variable by a 
factor of $\sim 2$, comparing the results from CRN99 with our data. 
This can be taken as indication for strong chromospheric activity, i.e. young age. 
Furthermore, Cha H$\alpha$ 1, 3, and 6, classified here as bona-fide BDs, are 
X-ray sources (NC98), an indication for magnetic activity, i.e. youth.

Comparing $W_{\lambda}$(Li) of our candidate and bona-fide BDs with TTS (Fig. 3) 
shows that the lithium strength seems to drop after M5 (B 34, $\log T_{eff} = 3.5$) 
and to increase again from $\sim$ M7 to M8. 
However, objects with $\le 0.3~M_{\odot}$, i.e. $\log T_{eff} \le 3.5$,
burn less than $1\%$ of their initial lithium in the first $\sim$ 10 Myrs,
so that the apparent gap may be due either to missing objects
or to an effect of molecular lines (eg. TiO in late M-type objects)
on the lithium curves-of-growth, i.e. on how the lithium abundance 
depends on $T_{eff}$ and $W_{\lambda}$(Li), so that the 
lithium-iso-abundance lines do not increase monotonically anymore
in the $W_{\lambda}$(Li) versus $T_{eff}$ plot, 
as they do for G- and K-type stars (see Fig. 3).
Moreover, the similar positions of our objects and of the Pleiades brown 
dwarfs (Basri et al. 1996, Rebolo et al. 1996, Stauffer et al. 1998) in the 
$T_{eff}$ versus $W_{\lambda}$(Li) diagram, despite a difference of more than 
one order of magnitude in age between both populations, seem to rule out 
evolutionary factors as the cause of the apparent gap.

Any object with detected lithium is young or sub-stellar or both. 
However, for any specific age (or age range or upper age limit), 
any object later than some specific spectral type is a BD. 
For the Pleiades (125 Myrs), Mart\'\i n et al. (1996) and Stauffer et al. (1998) 
found that any member later than M6.5 is a BD. 
One can also see from evolutionary tracks and isochrones that any
object with an age of $\le 125$~Myrs which is
cooler than the temperature corresponding to M6.5 lies below the hydrogen burning
limit (D'Antona \& Mazzitelli 1997, Burrows et al. 1997, Baraffe et al. 1998).

\begin{figure*}
\vbox{\psfig{figure=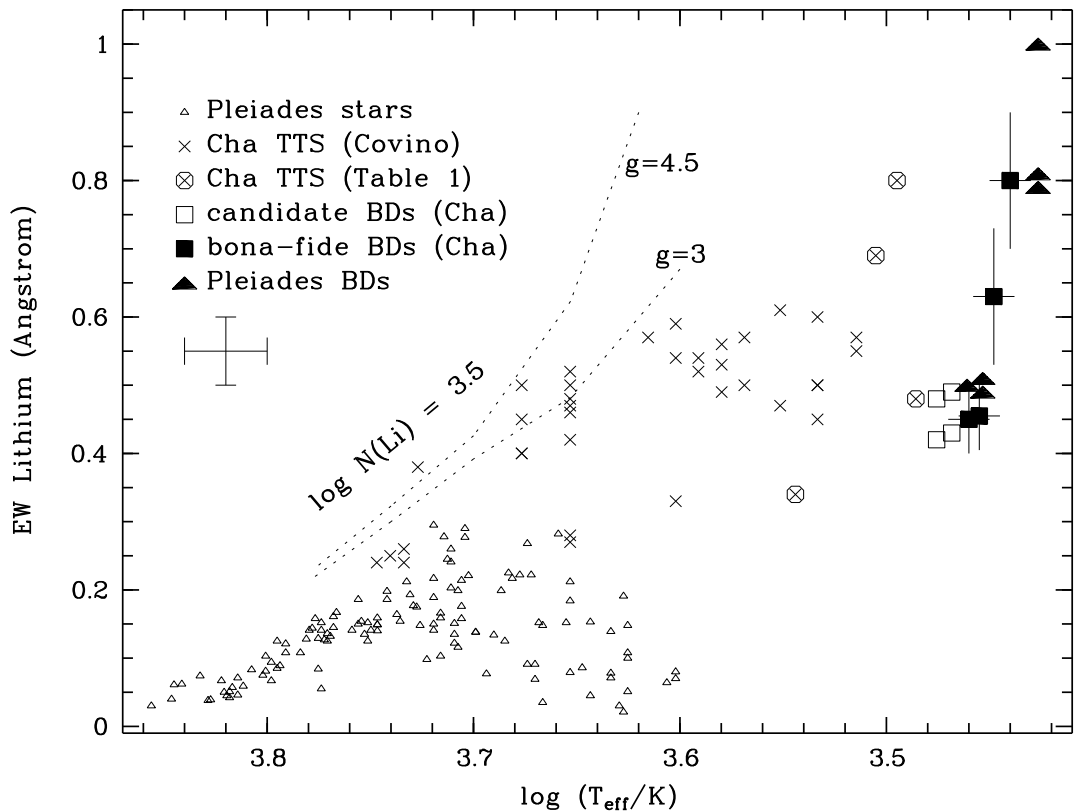,width=19cm,height=13cm,angle=270}}
\caption{ Equivalent width $W_{\lambda}$(Li) versus effective temperature $T_{eff}$
for Pleiades stars (Soderblom et al. 1993), Pleiades BDs (Basri et al. 1996, 
Rebolo et al. 1996, Stauffer et al. 1998), TTS in Cha (from Covino et al. 1997
and our Table 1), and Cha I bona-fide and candidate BDs. The lines indicate the
primordial lithium $\log$ N(Li)$=3.5$ for sub-giants ($\log g = 3.0$)
and dwarfs ($\log g = 4.5$) from Pavlenko \& Magazz\`u (1996) fitting
well as upper envelope to the G- and K-type TTS; lithium curves-of-growth 
for M-type objects are not yet available. 
For the F- to K-type TTS and Pleiades, we have converted spectral types to $T_{eff}$ 
using Bessell (1979, 1991), for the M-type objects using Fig. 7 in Luhman (1999) for 
objects intermediate between dwarfs and giants, taking into account that Pleiades
BDs are slightly colder than Cha I BDs at any given spectral type (Luhman et al. 1997)
}
\end{figure*}

Cha I members are younger than the Pleiades, i.e. lie higher above the main sequence, 
so that they are slightly hotter at any given spectral type (Luhman et al. 1997),
corresponding to roughly $\pm 0.25$ sub-classes. Considering also 
$\pm 0.25$ error in assigning the spectral classes, we should be sure that any 
Cha I member with spectral type M7 (i.e. $T_{eff} = 2890$ K, Luhman 1999) or later
is sub-stellar. This is the case for Cha~H$\alpha$~1, 3, 6, and 7, classified 
as bona-fide BDs in Table 1. This classification is correct, regardless of whether
the objects are young: either they are younger than $\sim$ 125 Myrs
and, hence, lie below the sub-stellar limit in the H-R diagram 
because of the late spectral type
and the young age (see tracks and isochrones by
D'Antona \& Mazzitelli 1997, Burrows et al. 1997, Baraffe et al. 1998);
or, alternatively, if they are not young, then they are sub-stellar because of the lithium. 
Actualy, because most Cha I TTS (see Lawson et al. 1996) as well as Cha~H$\alpha$~1 to 6 
(see CRN99) are $\sim 1$~Myrs old, all our bona-fide and candidate BDs are probably of 
that young age, too, i.e. much younger than $\sim 125$~Myrs.
Given the spectral types of Cha~H$\alpha$~2, 4, 5, and 8, they
are located too close to the limit for definitive discrimination. 

\acknowledgements{
We would like to thank Olivier Hainaut, Stephane Brillant and the NTT staff 
for the perfect support, Eduardo Mart\'\i n 
and Isabelle Baraffe for very useful discussion, 
Adam Burrows for providing us with unpublished lithium depletion calculation, 
and Laurent Cambr\'esy with the whole DENIS team 
for contributing unpublished DENIS data. }

{}

\end{document}